# A new data assimilation method of recovering turbulent flow field at high-Reynolds numbers for turbulence machine learning

Yilang Liu, Weiwei Zhang[*]

( School of Aeronautics, Northwestern Polytechnical University, No. 127 Youyi West Road, Xi'an 710072, China )

**Abstract:** This paper proposes a new data assimilation method for recovering high fidelity turbulent flow field around airfoil at high Reynolds numbers based on experimental data, which is called Proper Orthogonal Decomposition Inversion (POD-Inversion) data assimilation method. Aiming at the flows including shock wave discontinuities or separated flows at high angle of attack, the proposed method can reconstruct high-fidelity turbulent flow field combining with experimental distributed force coefficients. We firstly perform the POD analysis to the turbulent eddy viscosity fields computed by SA model and obtain the base POD modes. Then optimized the POD coefficients by global optimization algorithm coupling with the Navier-Stokes equations solver. The high-fidelity turbulent flied are recovered by several main modes, which can dramatically reduce the dimensions of the system. The effectiveness of the method is verified by the cases of transonic flow around the RAE2822 airfoil at high Reynolds numbers and the separated flow at high angles of attack. The results demonstrate that the proposed assimilation method can recover the turbulent flow field which optimally match the experimental data, and significantly reduce the error of pressure coefficients. The proposed data assimilation method can offer high-fidelity field data for turbulent model based on machine learning.

## 1. Introduction

Turbulence is considered as the last major unsolved problem in classical physics[1]. Numerical simulation methods play an increasingly important role in the research of turbulent problems. There are mainly three numerical methods to solve the Navier-Stokes equations: direct numerical simulation (DNS), large eddy simulation (LES) and Reynolds Navier-Stokes equation simulation (RANS). DNS and LES can resolve elaborate and smaller scales of turbulence, while both of the methods are expensive in computational cost, and can be only used for low Reynolds number flows currently. RANS method only resolves large scales of turbulence, which has high computational efficiency and costs less memory. Therefore, in the near future, RANS is still the dominant tool for dealing with the industrial problems.

*Corresponding author:
Weiwei Zhang: aeroelastic@nwpu.edu.cn

The traditional RANS model can be mainly divided into two categories: the first-order closures based on the Boussinesq hypothesis and the second-order closures that is the so-called Reynolds stress model. The first-order closures include algebraic model, such as Cebeci-Smith(CS) model and Baldwin-Lomax(BL) model[2], and one/two-equation model, such as the famous Spalart-Allmaras(SA) model[3], k-ε/k-ωmodel[4]~[5] and Menter's SST model[9]~[10]. Although the first-order closures have been widely used in engineering, they only suit for the simulation of the attached flows with low angles of attack. While, for the separated flows dominated by vortex, the accuracy of the first-order closures is often unreliable. The second-order closures directly constructed the exact equations for Reynolds stresses, which improved the accuracy for simulating turbulent flows. However, the model needs to establish the transport equations for the Reynolds stresses with six components, and it also has higher-order stress terms to be closed because of the nonlinearity of NS equations. Therefore, the Reynolds stress model is computationally expensive and less robust than the first-order closures. As a result, it has not been widely used in aerospace science and engineering.

The key difficulty that the traditional RANS models encounter is that they cannot compute the distributions of eddy viscosity or Reynolds stress tensors accurately for the separated turbulent flows, although they have the strong ability to simulate the attached flows. Therefore, in recent years, researchers developed many data-driven turbulent models based on the machine learning techniques. The main idea of these methods is that the high-fidelity data obtained by DNS or LES are analyzed with the assistance of the powerful computer and the advance machine learning technology, which are used to augment the RANS models. Milano et al[11] developed a neural network methodology to reconstruct the near wall turbulent channel flow by DNS data. Hocevar et al[12] constructed a radial basis function neural network to estimate the turbulent wake behind an airfoil based on the flow visualization. Duraisamy and Singh et al[13]~[16] modified the original SA model equation by introducing a multiplier of the production term, and developed the model for the extra multiplier combined with high-fidelity data. The augmented model was shown the significant improvement to the baseline model. Xiao et al[17]~[19] proposed a data-driven physics-informed machine learning approach for reconstructing the discrepancy between RANS modeled Reynolds stresses and DNS databases. The discrepancy model was used to modify the original RANS model, which also improve the accuracy of the turbulent simulation. Ling et al[20]~[21] proposed a Tensor Basis Neural Networks (TBNN) model to learn the Reynolds stress anisotropy tensor from high-fidelity DNS data. The TBNN model embedded Galilean invariance into the network and was shown more accurate predictions than the traditional RANS model. Marques et al[22] constructed a data-driven model for the boundary-layer profiles of the low Reynolds number airfoils, which was proved to be capable of producing smooth estimates of boundary-layer quantities. Zhu&Zhang et al[23] proposed a neural network to replace the original partial differential equation turbulent model. They directly reconstructed the relations between the turbulent viscosity and the mean flow variables. The model was coupled with NS solver and was applied to the high Reynolds number turbulent flow around the airfoils, which demonstrated excellent agreement with the training data of the SA model.

Although the recently developed machine learning turbulence models have made the remarkable achievements for turbulence simulation, these kinds of methods need to be based on a large number of fine and high-fidelity flow field samples. Due to the limited methods of the experimental measurement, the detailed information for turbulent flow field can not be easily obtained, especially the turbulent velocity, pressure field in the boundary layer and the friction coefficients. Therefore, such flow field samples can only be obtained by high-fidelity numerical methods(DNS/LES). Since it is still expensive for acquiring accurate turbulent flow field at high Reynolds numbers by DNS/LES methods, most of the current data-driven turbulent models can only be used for the flows with simple configurations at low Reynolds numbers. The acquisition of high-fidelity flow field samples is the bottleneck of the current machine learning methods for applying to the high Reynolds number flows.

Given the above problems, the data assimilation method has been used to obtain the high-fidelity flow field data and to improve the accuracy of the traditional RANS models. The main idea of these methods is that a small number of experimental data is taken as the high-fidelity reference data, and the numerical methods are used to compute the turbulent flowfield, then the data assimilation method are applied to modify the numerical models based on the high-fidelity data in order to minimize the discrepancy between the computational results and the experimental data. Foures et al[24] proposed a data assimilation method based on the variational formulation and the Lagrange multipliers approach to reconstruct the full mean flow field. They considered the RANS Reynolds stresses as an external forcing term, and the data assimilation method was used to determine the appropriate term which can minimize the error between the DNS data and the numerical solution of RANS equations. Symon et al[25]~[26] extended the data assimilation method based on Foures' work to reconstruct mean flow at relatively higher Reynolds number flows combining with the experimental velocity, and they also introduced the resolvent analysis to assist in selecting experimental reference points. Kato et al[27] applied the data assimilation method based on the ensemble transform Kalman filter to estimate and correct the initial conditions of wind tunnel, such as the angles of attack and Mach numbers, and they also revised the turbulent viscosities for RANS model and improved the accuracy of the current turbulence models. In addition, the data assimilation methods have also been used to modify the empirical parameters in traditional RANS model equations in order to improve the numerical accuracy of turbulence flows according to high-fidelity data. Kato et al[28] employed the ensemble Kalman filter to investigate the parameters in SA model with the simulation of flow around a flat plate. Li&Martin et al[29] developed a data-driven adaptive RANS k-ω model, which can automatically adapt the k-ω closure coefficients to improve agreement with the experimental data compared with the original RANS k-ω model. Deng&Liu et al[30] adopted the ensemble-Kalman-filter-based data assimilation method to recover the global flow field, and they calibrated the empirical parameters for four different RANS models according to the experimental measurement data. The significant reduction in spatial-averaged error distribution can be achieved by the modified models. Moreover, the data assimilation method have also been applied for recovering unsteady viscous flows[31] and large eddy viscosity simulations[32]~[33].

Most of the existing RANS turbulent models, including the type of traditional partial differential equations and the recently developed machine learning turbulent models, mainly concentrate on the attached flows or low Reynolds number flows. The current data assimilation method is one of the effective ways to obtain high-fidelity turbulent flow data at high Reynolds numbers quickly and accurately. However, most of the existing data assimilation methods concentrate on the estimation for the model parameters, while it is expensive for them to directly reconstruct the whole flow field. Since the turbulent flow at high Reynolds numbers must have tremendous degrees of freedom, the current data assimilation methods often lead to an extremely high-dimensional optimization problem. Therefore, it is difficult to find out the global optimal solution. Aiming at dealing with the problems for high Reynolds number flows around airfoils in engineering, including flows with shock waves or separation at high angles of attack, this paper proposes a new POD-Inversion data assimilation method for recovering high-fidelity turbulent field which can optimally match the experimental data. The proposed method firstly performs modal analysis for the samples of turbulent eddy viscosity field by the POD technique and constructs the basic POD modes. Then, we select the first few main modes and optimize the modal coefficients by using global optimization algorithm. The POD-Inversion method can dramatically reduce the dimension of flow variables, and conveniently reconstruct the high-fidelity turbulent viscosity flow fields which conform more accurately with the experimental data. This work will provide an efficient way to obtain the high-fidelity turbulent flows in order to lay the foundation for machine learning turbulence models at high-Reynolds numbers.

The outline of this paper is as follows. Section 2 describes the CFD governing equations and the methodology of the POD-Inversion method. Section 3 presents the numerical examples to verify the proposed method, and the conclusions are drawn in section 4.

## 2. Methodology

### 2.1 Governing equations

The integral form of two-dimensional Reynolds average Navier-Stokes equations can be written as:

$$\frac{\partial}{\partial t}\iint_\Omega Q d\Omega + \oint_{\partial\Omega} F(Q)\cdot n d\Gamma = \oint_{\partial\Omega} G(Q)\cdot n d\Gamma \qquad 1$$

where $\Omega$ is the control volume; $\partial\Omega$ is the boundary of control volume; and $n=(n_x,n_y)^T$ denotes the unit outward normal vector to the boundary. The vector of conservative variables $Q$, inviscid fluxes $F(Q)=(F_x(Q),F_y(Q))$ and viscous fluxes $G(Q)=(G_x(Q),G_y(Q))$ are given as follows:

$$Q = \begin{Bmatrix} \rho \\ \rho u \\ \rho v \\ E \end{Bmatrix} \qquad F_x(Q) = \begin{Bmatrix} \rho u \\ \rho u^2 + p \\ \rho uv \\ u(E+p) \end{Bmatrix} \qquad F_y(Q) = \begin{Bmatrix} \rho v \\ \rho uv \\ \rho v^2 + p \\ v(E+p) \end{Bmatrix} \qquad 2$$

$$G_x(Q) = \begin{Bmatrix} 0 \\ \tau_{xx} \\ \tau_{xy} \\ u\tau_{xx}+v\tau_{xy}-q_x \end{Bmatrix} \qquad G_y(Q) = \begin{Bmatrix} 0 \\ \tau_{yx} \\ \tau_{yy} \\ u\tau_{yx}+v\tau_{yy}-q_y \end{Bmatrix} \qquad 3$$

where the viscous stresses and the heat fluxes are

$$\tau_{xx} = 2(\mu+\mu_T)\left[\frac{\partial u}{\partial x} - \frac{1}{3}(\frac{\partial u}{\partial x}+\frac{\partial v}{\partial y})\right]$$

$$\tau_{yy} = 2(\mu+\mu_T)\left[\frac{\partial v}{\partial y} - \frac{1}{3}(\frac{\partial u}{\partial x}+\frac{\partial v}{\partial y})\right] \quad 4$$

$$\tau_{xy} = \tau_{yx} = (\mu+\mu_T)(\frac{\partial u}{\partial y}+\frac{\partial v}{\partial x})$$

$$q_x = -\frac{1}{(\gamma-1)}(\frac{\mu}{\Pr}+\frac{\mu_T}{\Pr_T})\frac{\partial T}{\partial x}$$

$$q_y = -\frac{1}{(\gamma-1)}(\frac{\mu}{\Pr}+\frac{\mu_T}{\Pr_T})\frac{\partial T}{\partial y} \quad 5$$

where $\rho$ denotes the density; $u$ and $v$ are the $x$ and $y$ direction components of the velocity vector; $p$ is the pressure; $E$ is the total energy per unite volume; $\mu$ and $\mu_T$ are the dynamic molecular viscosity and the turbulent eddy viscosity respectively; $T$ is the temperature; $\Pr$ and $\Pr_T$ are the laminar and turbulent Prandtl number; and $\gamma$ is the ratio of specific heats. For the ideal gas, $\gamma$ is equal to 1.4. According to the Sutherland's law, the dynamic viscosity coefficient is given by

$$\mu = \mu_{ref}\frac{T_{ref}+S_0}{T+S_0}(\frac{T}{T_{ref}})^{\frac{3}{2}} \quad 6$$

where $T_{ref}$ and $\mu_{ref}$ are physical constants of reference temperature and viscosity, and $S_0$ is the Sutherland temperature. The values of them are $T_{ref}=273.15K$, $\mu_{ref}=1.716\times10^{-5}kg/(m\cdot s)$ and $S_0=110K$, respectively. The state equations for the ideal gas is

$$p = (\gamma-1)[E-\frac{\rho}{2}(u^2+v^2)] \quad 7$$

The viscous stresses satisfy the equation (4) in condition of Boussinesq hypothesis, which assumes that the turbulent shear stress depends linearly on the mean rate of strain. The turbulent eddy viscosity coefficient $\mu_T$ need to be solved by the closure models, such as the commonly used one-equation Spalart-Allmaras (SA) turbulence model and two-equation $K-\varepsilon$ SST (Shear-Stress Transport) model. The turbulent eddy viscosity for SA and $K-\varepsilon$ SST models can be written as follows,

$$\mu_T = f_{v1}\rho\tilde{\upsilon} \quad 8$$

$$\mu_T = C_\mu f_\mu \rho\frac{K^2}{\varepsilon} \quad 9$$

where $\tilde{\upsilon}$ and $f_{v1}$ are the eddy viscosity variable and model constant for SA model respectively. $K$ and $\varepsilon$ denote the turbulent kinetic energy and dissipation rate, and $C_\mu$ and $f_\mu$ are model constants for SST model.

It is obviously that the key task of the first-order closures for RANS models is to compute the term of the turbulent eddy viscosity. Both the above models need to construct additional partial differential equations for turbulent variables to close the RANS equations. Since the additional equation, such as SA model, were developed based on empiricism, dimensional analysis and Galilean invariance, they are actually the semi-theoretical-semi-empirical models which contain many empirical model constants. Therefore, the traditional RANS models can only suit for the attached turbulent flows, and they usually can not give the satisfactory results for flows containing shock waves and separations. One of the significant reasons is that the turbulent eddy viscosity can not be solved accurately.

## 2.2 POD-Inversion data assimilation method

Most of the recent studies concentrated on calibrating the parameters[27]~[29], [30] or adding additional correction items[15]~[16], [34] in traditional turbulent models based on the framework of partial differential equations. Data assimilation methods, especially by Kalman filtering technique[24]~[28], [30]~[33], are adopted to improve the accuracy of RANS simulation. The KF combines the input state vectors with the experimental observation values, usually the velocity or pressure field, to make an optimal estimation for the system variables. The reference [32] also found that the number of ensembles of data assimilation method has big effects on the state estimation, and decreasing the number of ensembles has a negative impact on the precision of estimation. As for flows at high-Reynolds numbers, the flow variables have tremendous degrees of freedom, so enough ensembles must be used to ensure the accuracy of recovering turbulence. Therefore, it will lead to a great high-dimensional optimization problem, which is extremely expensive and difficult to find the optimal solutions.

In this paper, we firstly implement the modal analysis by POD technique to the turbulent eddy viscosity flow field, and choose the first few main POD modes as the basic modes to represent the complex field at high-Reynolds number. It can dramatically reduces the dimension of the systems, and recovers high fidelity turbulent flow field efficiently and conveniently.

The POD technique originated from the principal component analysis method in statistics, which has powerful and effective ability for data dimensionality reduction analysis. The most advantage of POD is that it can project high-order, high-dimensional and non-linear systems onto a low-dimensional state space through orthogonal modals, and meanwhile maintains the minimum residual error in a given number of modes. It was introduced to the fluid dynamics by Lumley[35] as a mathematical technique to extract coherent structures from turbulent flow fields. The POD has been developed to be one of the most widely used techniques in analyzing fluid flows, which applied to many different research areas, including data compression[36], reduced-order modeling[37], flow control[38] and aerodynamic design optimization[39].

Firstly, we compute the steady flows with SA model and obtain N turbulent eddy viscosity fields for different flow states(Mach numbers and angles of attack). We define the matrix of the turbulent eddy viscosity vector in equation (10),

$$\boldsymbol{X} = [\boldsymbol{X}_1(\mu_T), \boldsymbol{X}_2(\mu_T), \cdots, \boldsymbol{X}_N(\mu_T)], \qquad \boldsymbol{X} \in \mathbb{R}^{M \times N} \qquad (10)$$

where the column vectors $\boldsymbol{X}_i(\mu_T)$ denote the eddy viscosity on all of the grids for the different snapshots. *M* is the total number of grids, and *N* represents the number of the snapshots.

Then, define the correlation matrix,

$$\boldsymbol{C} = \boldsymbol{X}^T \boldsymbol{X}, \qquad \boldsymbol{C} \in \mathbb{R}^{N \times N} \qquad (11)$$

Since, in our problems, the spatial size *M* is much larger than the number of snapshots *N*, the correlation matrix will be changed to a much smaller and computationally more tractable eigenvalue problem. Solving the eigenvalue problem of the size *N*×*N* can easily find the POD modes,

$$\boldsymbol{C}\boldsymbol{\psi}_j = \lambda_j \boldsymbol{\psi}_j, \qquad \boldsymbol{\psi}_j \in \mathbb{R}^N \qquad (12)$$

Where $\lambda_j$ and $\psi_j$ are the eigenvalues and eigenvectors of the correlation matrix respectively. Therefore, we can obtain the POD modes $\Phi$ for the turbulent eddy viscosity field through

$$\Phi = X\Psi\Lambda^{-1/2} \qquad \Phi \in \mathbb{R}^{M \times N}$$
$$\Phi = [\phi_1, \phi_2 \cdots \phi_N]$$
$$\Psi = [\psi_1, \psi_2 \cdots \psi_N] \qquad (13)$$
$$\Lambda = [\lambda_1, \lambda_2 \cdots \lambda_N]$$

Since the POD modes are sorted according to the magnitude of energy, it is usually possible to accurately reconstruct the characteristics of the entire flow field by holding the leading main *r* modes. The recovering turbulent eddy viscosity flow field can be represented as

$$X_i(\mu_T) = \sum_{n=1}^{r} a_i^n \psi_i^n \qquad (14)$$

where the expansion coefficients $a_i^n$ can be determined according to

$$a_i = \langle X(\mu_T), \psi_i^n \rangle \qquad (15)$$

The POD modes and the expansion coefficients in equation (14) are obtained according to the samples of SA model. And in order to reconstruct the turbulent eddy viscosity distribution conformed to the experimental data, we should find a new set of POD coefficients. In view of this consideration, orienting to the experimental pressure distributions, we adopt the TLBO global optimization algorithm to optimize the modal coefficients. RANS equations are solved with the fixed the eddy viscosity in every iterative step, which are determined by the equation (14), and the optimal modal coefficients can be conformed when the optimization algorithm converges. The final turbulent eddy viscosity field can be expressed as

$$X_{opti}(\mu_T) = \sum_{n=1}^{r} a_{opti}^n \psi^n \qquad (16)$$

**Remark 1:** The significant feature of the POD-Inversion data assimilation method is to perform modal analysis to the high-dimensional turbulent eddy viscosity fields, and recover high-fidelity turbulent flow according to the non-linear base modes in low-dimensional space. The eddy viscosity is loosely coupled with the NS equations in the process of iterations, that is, we fix the eddy viscosity and directly assign it to the NS equations in every single step of optimization. And at the different iterative steps of optimization, the eddy viscosity is changed by variational POD expansion coefficients. Finally, the turbulent flow field we obtained is satisfied with NS equation, which ensure the validity of the POD-Inversion method.

**Remark 2:** Recovering the turbulent flow field only by limiting the distributed forces, such as the pressure or friction coefficients, can be considered as a mapping from the low-dimensional target to the high-dimensional solutions. Therefore, the turbulent flow field is not the unique solution to the RANS equations in a certain state, but is the converged solution which optimally matches the experimental data. Despite we can not achieve the unique solution, we will show in the test cases that the high-fidelity turbulent flow field can be recovered only by the limited pressure coefficients.

**Remark 3:** The POD-Inversion method needs to solve the steady flow field in every step of optimization. We stop the iteration in condition that the residual converges to $10^{-8}$ or the maximum iteration step reaches to 10000, and the turbulent flow fields whose residual value reaches to $10^{-5}$ at least will be selected as the candidate solutions in the process of optimization. We also find that it has good convergence property for the attached flows or the separation-fixed flows. While for flows near the stall point, the convergence for the loose coupling method shows not very well since the flow tends to be unsteady, and the stable solution can not be obtained by solving steady equations.

## 3. Numerical cases

### 3.1 Transonic flow past a RAE2822 airfoil

The transonic turbulent flow past a RAE2822 airfoil is firstly used to evaluate the performance of the developed POD-Inversion data assimilation method. The computational mesh is shown in Figure 1, which consists of 43325 elements and 400 boundary points on the airfoil surface. The free stream Mach number is $Ma = 0.729$, the angle of attack is $\alpha = 2.31°$ and the Reynolds number is $Re = 6.5 \times 10^6$. We fix the Reynolds number, and select the different states of angle of attack and the Mach number. The sample space is $\alpha \in [2°, 3°]$ and $Ma \in [0.72, 0.74]$, and we choose 10 samples by Latin hypercube method, which are shown in Figure 2. POD analysis is performed to the turbulent eddy viscosity flow fields obtained by SA model, and the leading four POD modes and coefficients are shown in Figure 3 and Figure 4.

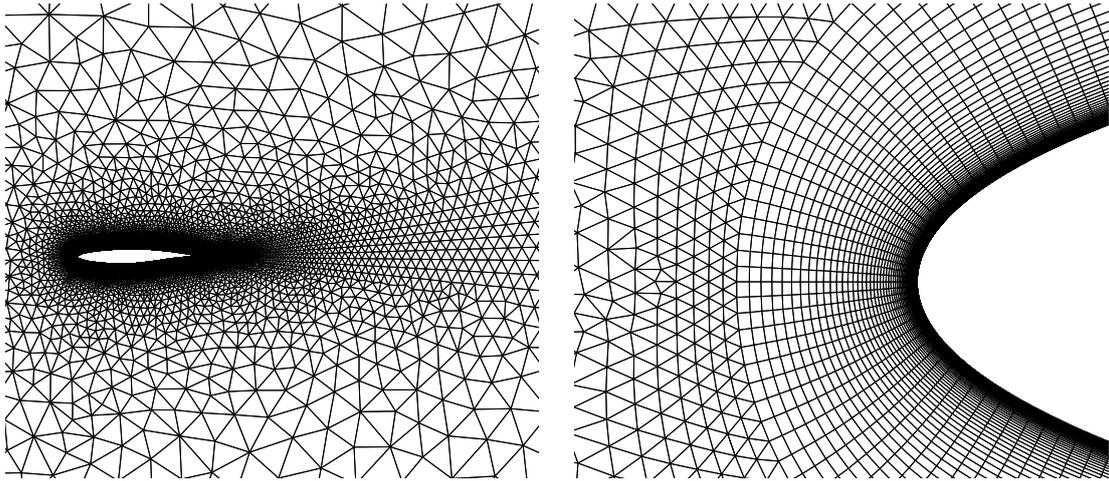

Figure 1 Grids near the RAE2822 airfoil

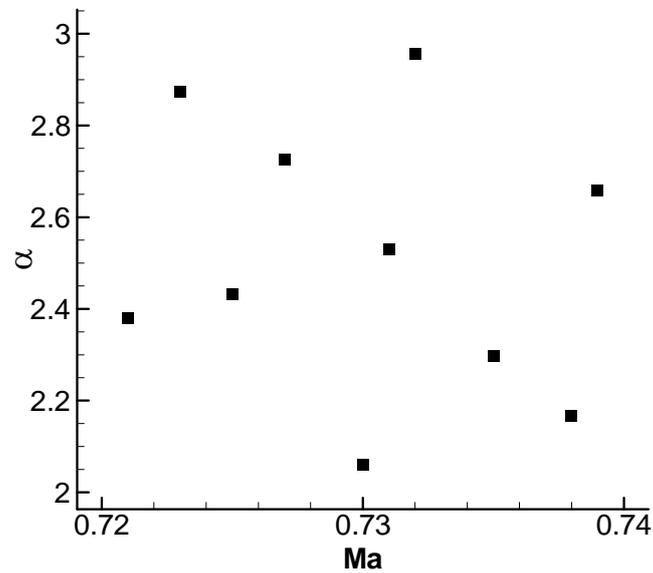

Figure 2 The samples for RAE2822 airfoil by Latin hypercube method

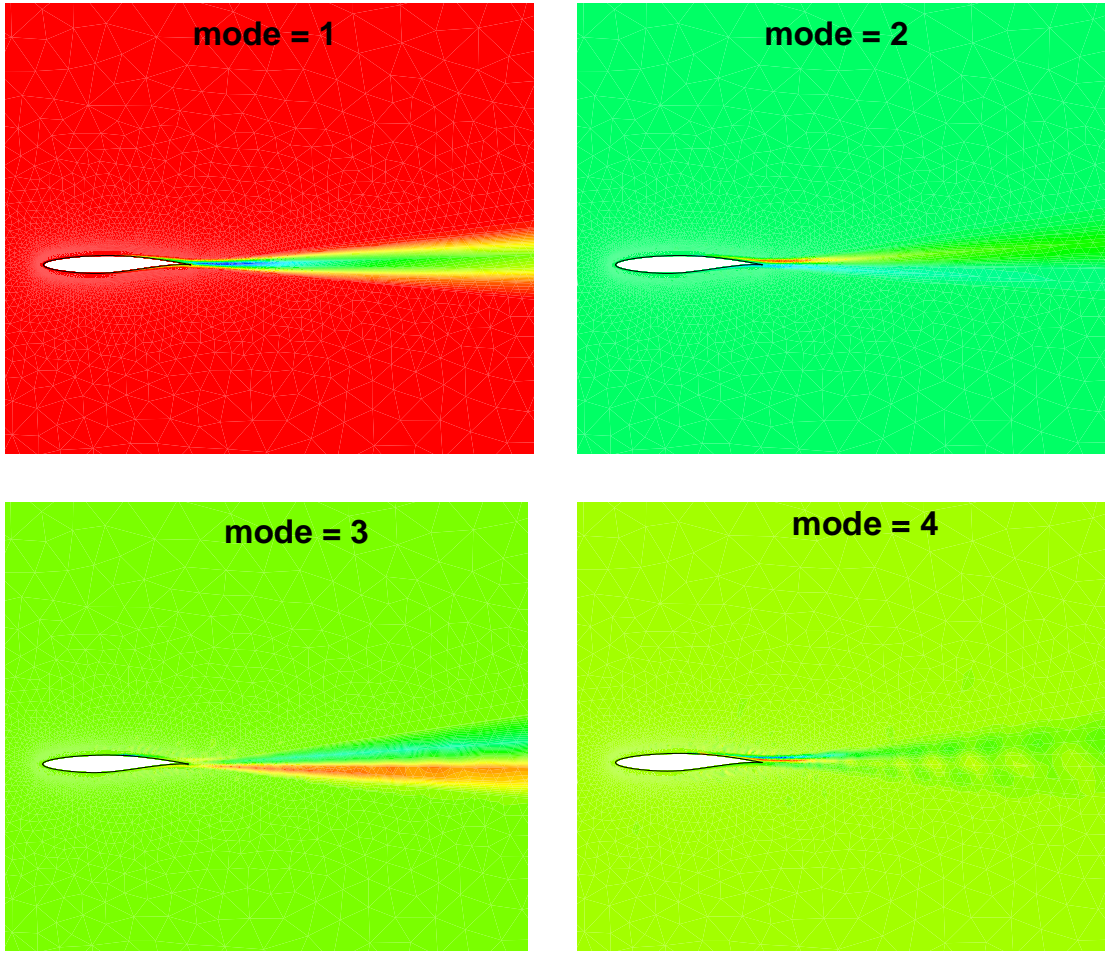

Figure 3 The leading four POD modes of turbulent eddy viscosity field for RAE2822 airfoil

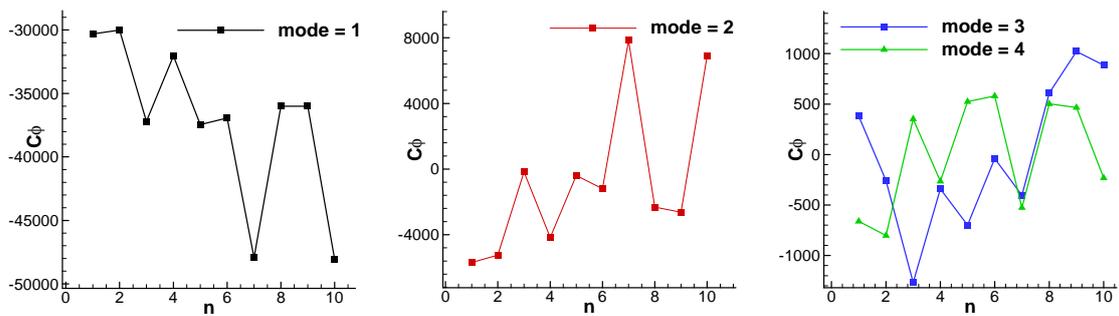

Figure 4 The leading four POD coefficients of turbulent eddy viscosity field for RAE2822 airfoil

The proposed POD-Inversion data assimilation method is used to recover the turbulent flow field. Since there is not enough experimental friction coefficients available, and we only take the experimental pressure coefficients into consideration, nevertheless the proposed method is internally consistent. Therefore, our ultimate goal is to reduce the error between the experimental and the simulated pressure coefficients as much as possible by adjusting the turbulent eddy viscosity flow field which is reconstructed by POD modes and expansion coefficients. We choose the first 8 POD modes as the base modes. TLBO algorithm is adopted to find the optimal modal coefficients, and the initial population is set as 20. The convergence history of TLBO is shown in Figure 5, where the x-axis represents the number of optimized generations and the y-axis denotes the error of pressure coefficients between the experimental and numerical data in terms of L2 norms, which is defined as:

$$Error = \frac{1}{N}\sqrt{\sum_{i=1}^{N}\left(C_{p\_exp}^{i} - C_{p\_cal}^{i}\right)^{2}} \qquad 17$$

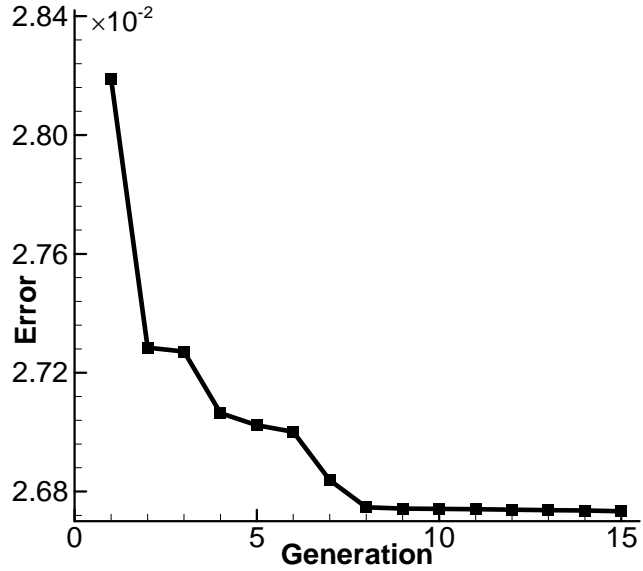

Figure 5 The convergence history of TLBO algorithm

It can be seen from the Figure 5 that the error remains decreased until the generation reaches to 8. The turbulent eddy viscosity after data assimilation and the discrepancy with those of SA model are shown in Figure 6. The results computed by POD-Inversion method is slightly smaller than those of the SA model, and the distributions are distinctly different near the RAE2822 airfoil and in the wake region, while the discrepancy is relatively small in other areas. The discrepancies of velocity in $x$-axis and pressure field between POD-Inversion method and SA model are displayed in Figure 7, and we can further see that the recovering flow field represents significant difference near the shock wave on the upper surface of the airfoil.

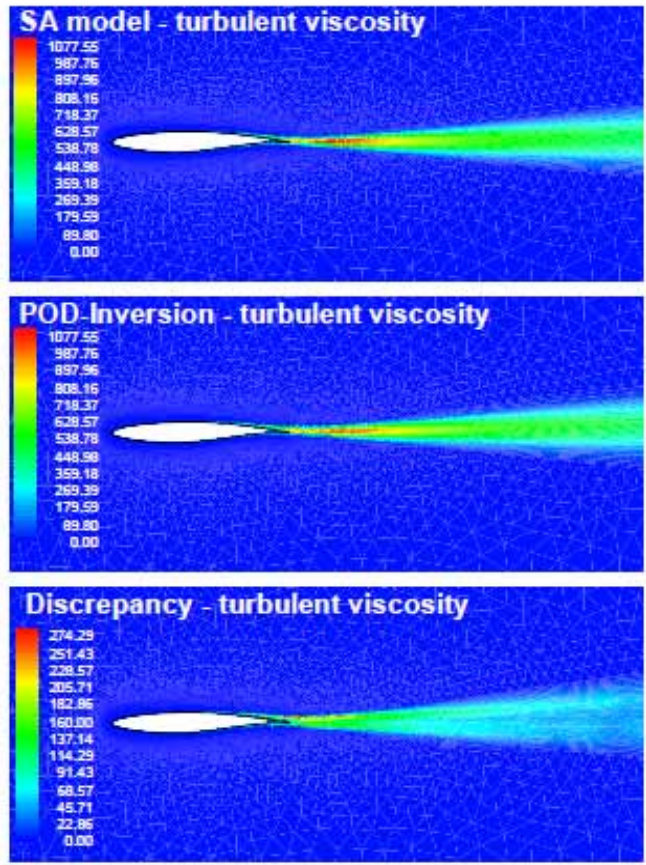

Figure 6 The comparisons of turbulent eddy viscosity flow field with POD-Inversion and the original SA model

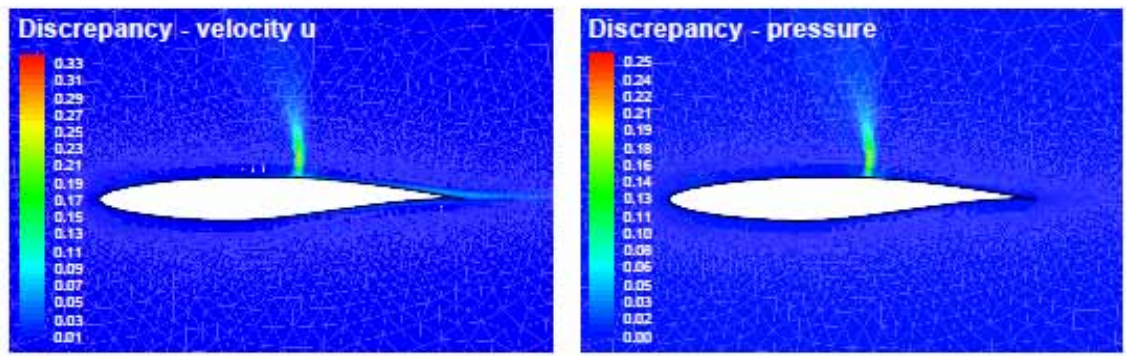

Figure 7 The dependency of velocity in *x*-axis and pressure distribution between POD-Inversion method and the original SA model

Figure 8 shows the comparisons of pressure coefficients between the both method and the experimental data[40]. The main difference between SA model and the experimental data of pressure coefficients for this case is the position of the shock wave, which can be ingeniously changed by slight modifying the flow field of the turbulent eddy viscosity. The results demonstrate that the POD-Inversion data assimilation method can accurately compute the position of the shock wave and the distributions of the pressure coefficients.

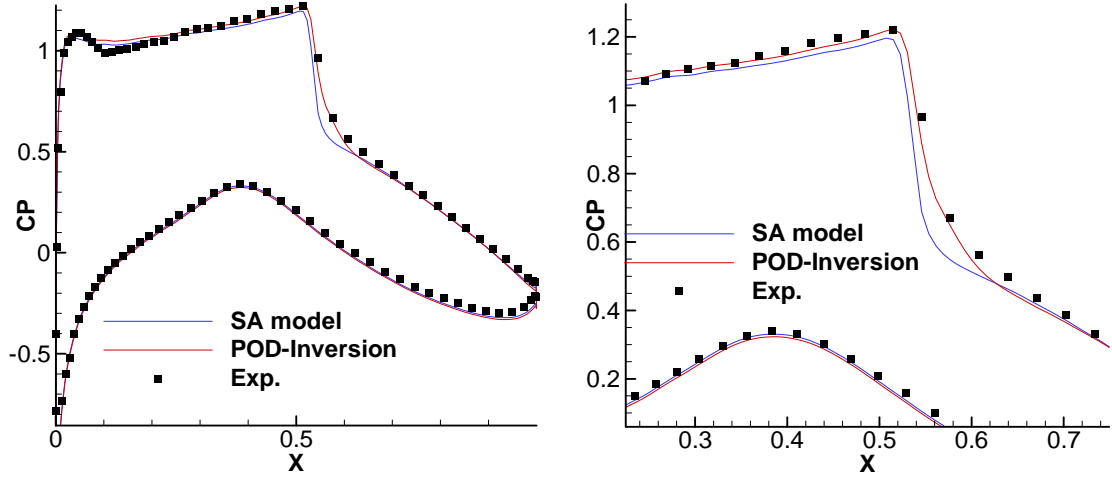

Figure 8 The pressure coefficients calculated by the POD-Inversion method and the original SA model with the comparison of experimental data

**3.2 The separated flow around a S809 airfoil at high angles of attack**

We further apply the proposed POD-Inversion data assimilation method for the S809 airfoil at high angles of attack. The computational mesh is shown in Figure 9, in which the total number of the elements is 36077, with 400 nodes on the airfoil surface. There are 40 layers of mesh in the boundary layer with the growth rate 1.1, and the first grid height is $8 \times 10^{-6}$. The free stream Mach number is $Ma = 0.15$, and the Reynolds number is $Re = 2 \times 10^6$.

We compute the turbulent flow fields of the S809 airfoil with SA model at the angle of attack from 0 to 15 degrees. The comparisons of lift coefficients with the angles of attack between SA model and experimental data[41] are displayed in Figure 10. For the traditional SA model, the computed lift coefficients are coincided with the experimental data in the condition of attached flows at the state of $\alpha \leq 8°$. However, when the angle of attack is larger than 8°, the results of SA model are dramatically different from the experimental data. This is mainly because SA model can not accurately simulate the separation point and region for the unattached flows.

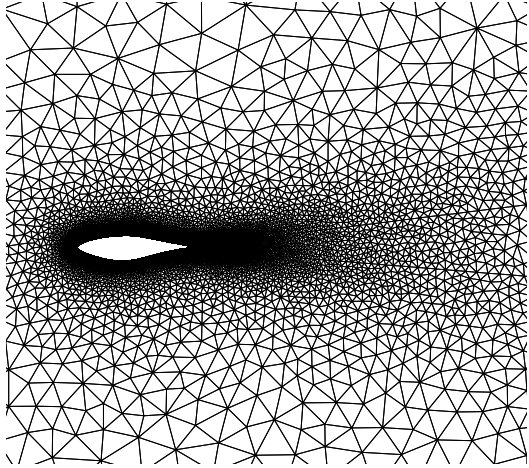 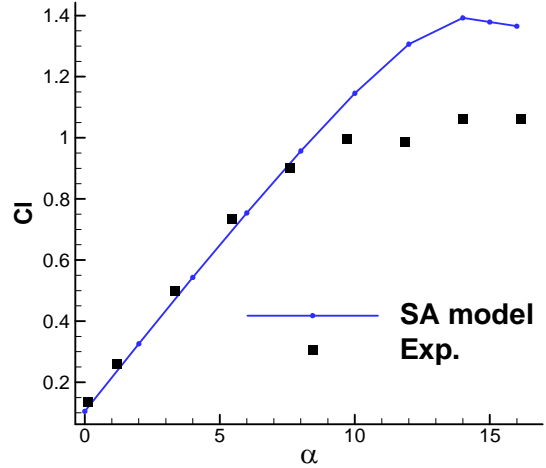

Figure 9 The computational mesh for S809 airfoil

Figure 10 The comparisons of lift coefficients with the angles of attack between SA model and experimental data

Therefore, concentrating on the state of $\alpha > 8°$, we adopt the POD-Inversion data assimilation method to recover the turbulent flow fields. Firstly, We also fix the Reynolds number, and select the different states of angle of attack and the Mach number. The sampling space is $\alpha \in [8°, 15°]$ and $Ma \in [0.1, 0.2]$, and we choose 10 samples for different flow states by Latin hypercube method, which are shown in Figure 11.

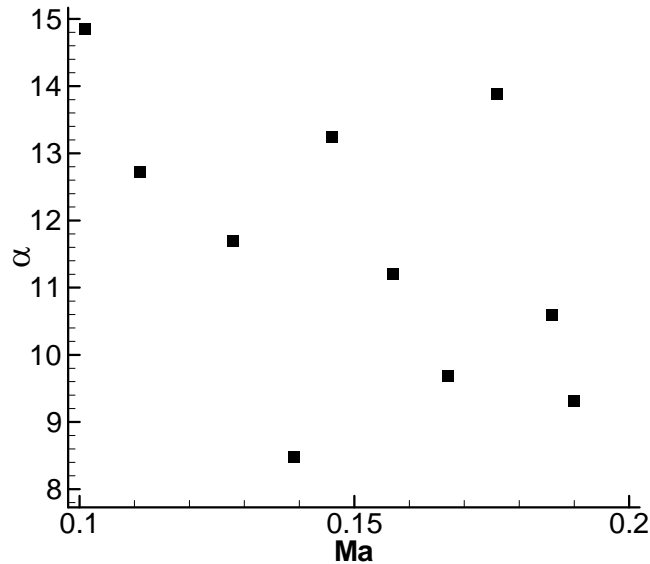

Figure 11 The samples for S809 airfoil by Latin hypercube method

The POD analysis is performed to the turbulent eddy viscosity flow field based on the samples computed by SA model, and the first four POD modes are shown in

Figure 12.

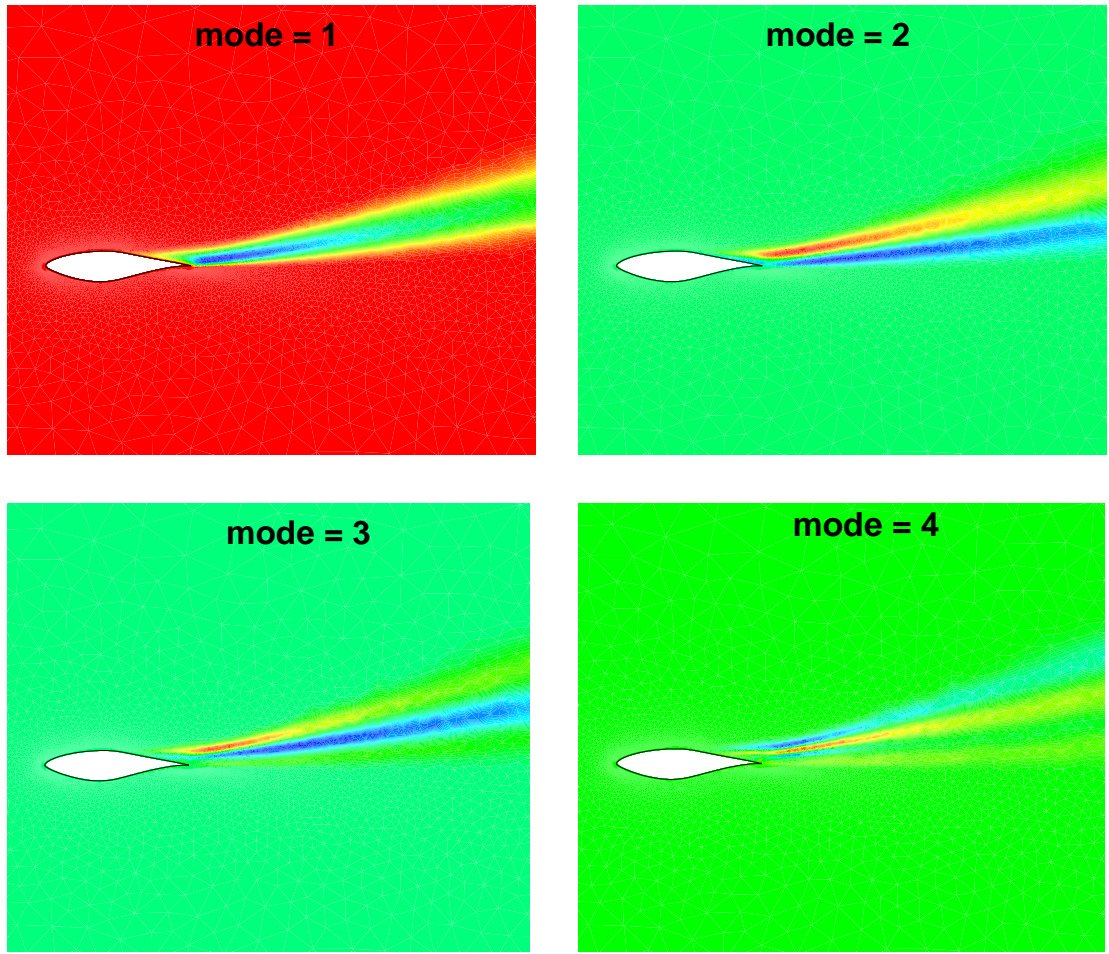

Figure 12 The leading four POD modes of turbulent eddy viscosity field for S809 airfoil

The leading 8 modes are remained to reconstruct the turbulent eddy viscosity flow field. The experimental data of the pressure coefficients are extracted from the reference [41], which are used as the optimization target. We mainly recover the turbulent flow field for the states of $\alpha = 8.2°$, $10.2°$, $12.2°$ and $14.2°$. The comparisons of velocity distributions and the streamlines for the POD-Inversion method and the SA model are shown in Figure 13. It can be seen from the results, SA model failed to calculate the separated region accurately at the trailing edge of the airfoil, and a tiny separated vortex just appears at the trailing edge when the angle of attack increases to $12.2°$. On the other hand, for the POD-Inversion method, an obvious separated vortex appears on the upper airfoil at the trailing edge when the angle of attack reaches to 10.2°. In addition, as the angle of attack increases, the area of the separation rapidly becomes larger, and at the state of $\alpha=14.2°$, the separation point calculated by POD-Inversion method is much closer to the leading edge than the SA model, and the separated region is also relatively larger.

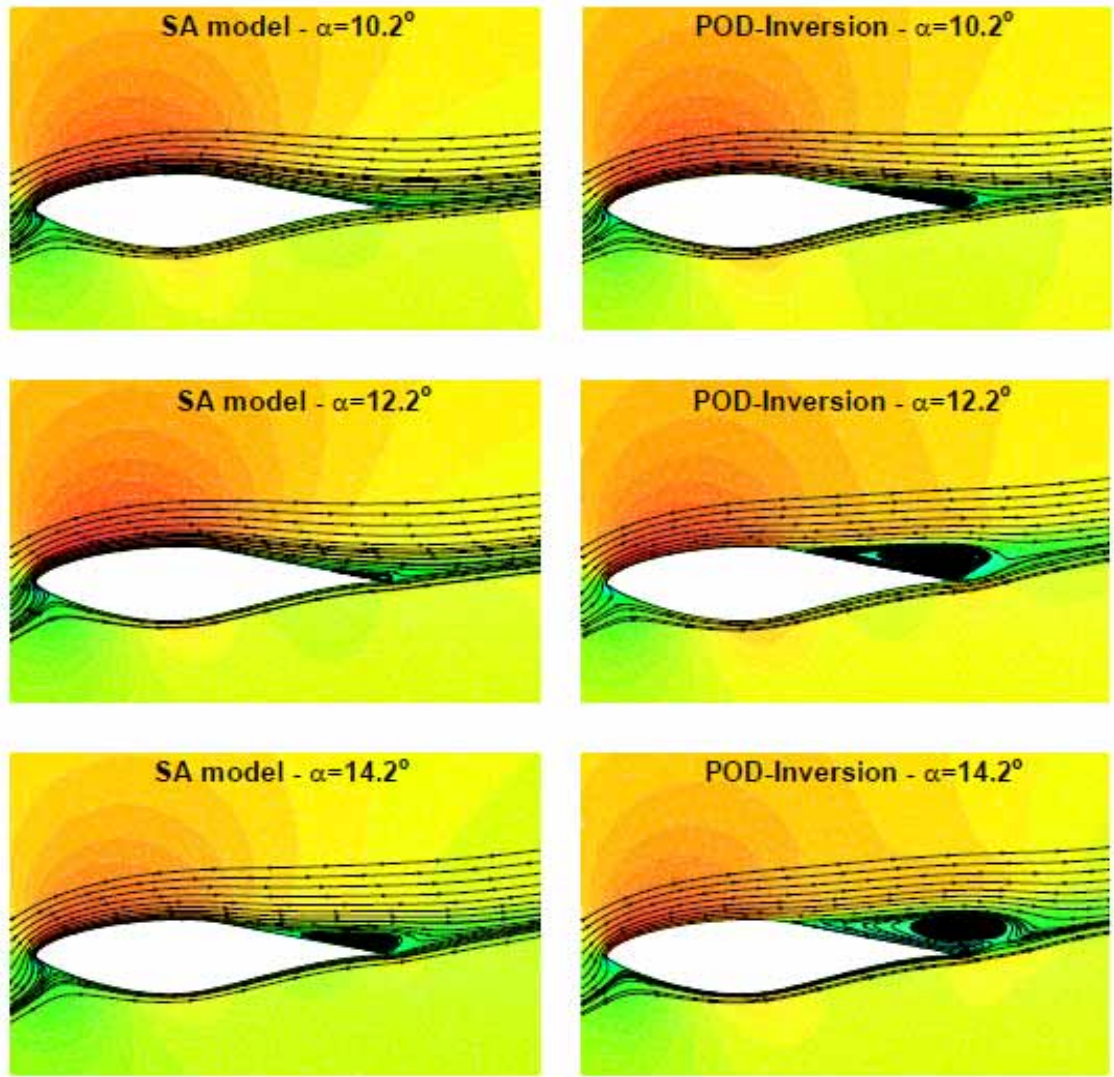

Figure 13 The comparisons of velocity distributions and the streamlines for the POD-Inversion method and the SA model

Figure 14 displays the distributions of turbulent eddy viscosity at the angle of 12.2° for the both methods. The value of eddy viscosity for POD-Inversion method, especially at the separated region, is relatively larger than SA model. Therefore, the increase of viscosity promotes the separation at the trailing edge and the large separation area.

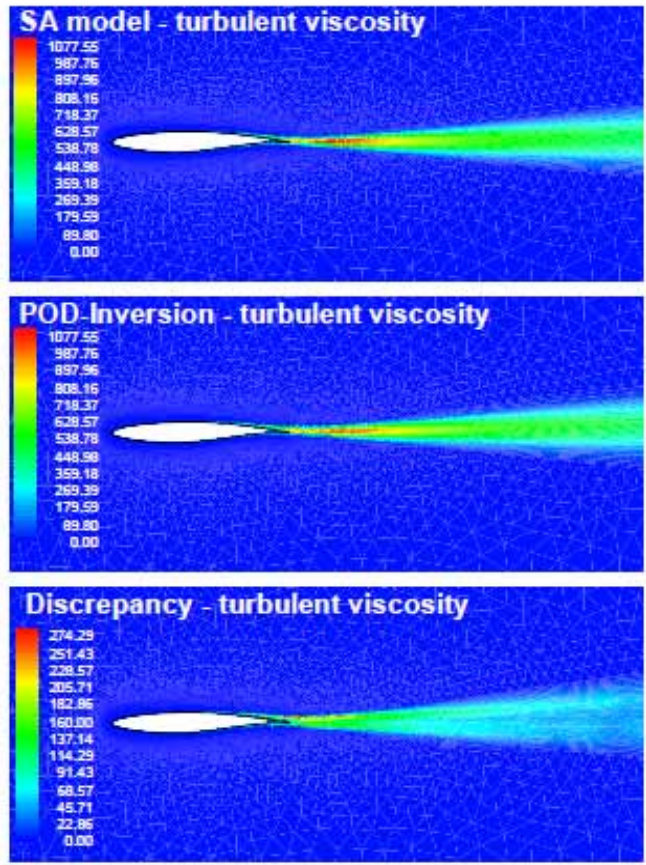

Figure 6 The comparisons of turbulent eddy viscosity flow field with POD-Inversion and the original SA model

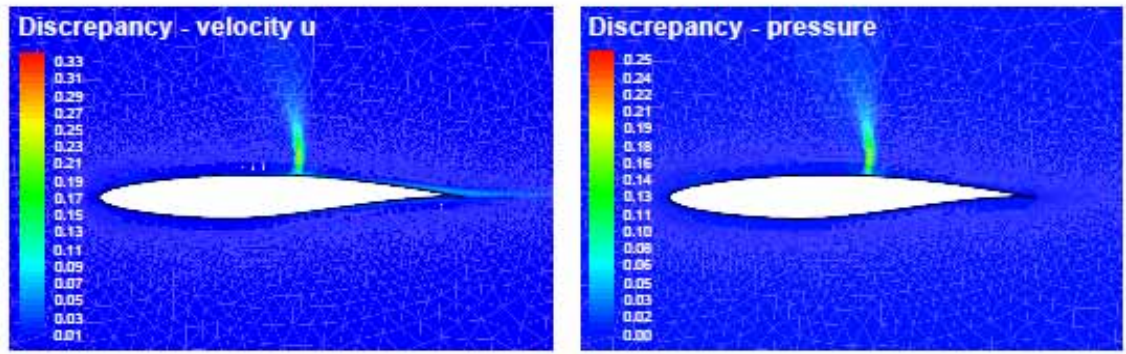

Figure 7 The dependency of velocity in *x*-axis and pressure distribution between POD-Inversion method and the original SA model

Figure 8 shows the comparisons of pressure coefficients between the both method and the experimental data[40]. The main difference between SA model and the experimental data of pressure coefficients for this case is the position of the shock wave, which can be ingeniously changed by slight modifying the flow field of the turbulent eddy viscosity. The results demonstrate that the POD-Inversion data assimilation method can accurately compute the position of the shock wave and the

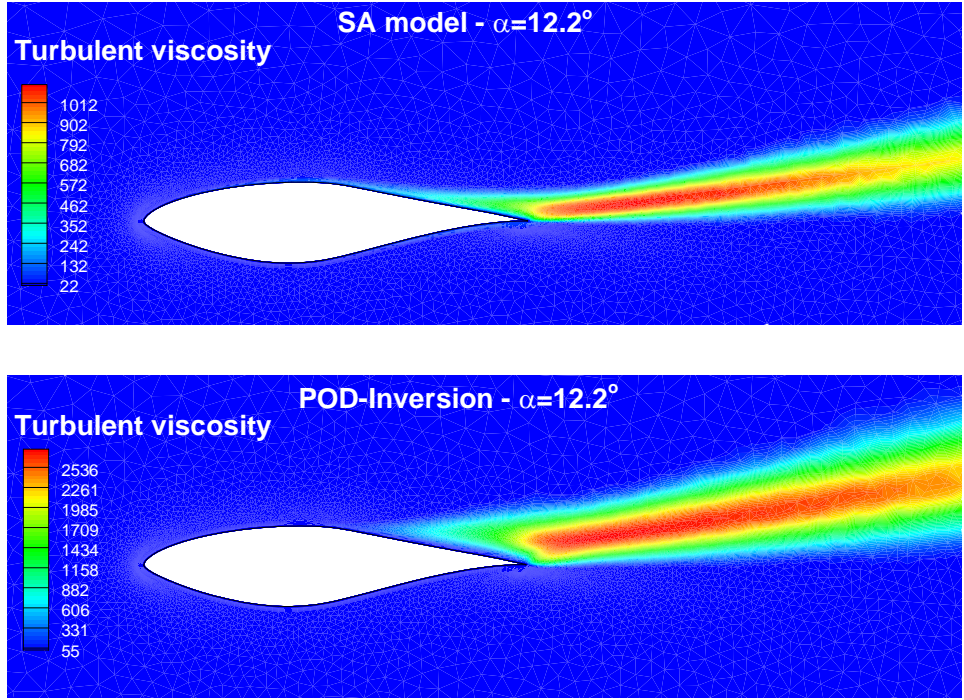

Figure 14 The distributions of turbulent eddy viscosity at $\alpha = 12.2°$ for SA model and POD-Inversion method

The comparisons of the pressure coefficients calculated by the two method with the experimental data for the four different flow states are shown in Figure 15. Here, we can see from the results, when the flow does not separate at the state of $\alpha = 8.2°$, the pressure coefficients computed by the both methods can be consistent with the experimental data. It indicates that the traditional SA model has the ability to accurately simulate the attached flows. However, when the angle of attack is further increasing, the separation emerges at the trailing edge on the upper surface of the airfoil. The discrepancy of pressure coefficients between SA model and the experimental data becomes distinct on the upper surface. Whereas, the results calculated by the POD-Inversion method can optimal coincide with the experimental data for all the flow states. Figure 16 shows the comparisons of lift coefficients versus angle of attack between experimental data and numerical results (at the angle $\alpha > 8.2°$, there is no need to perform data assimilation method since SA model can give the accurate results, so we consider the POD-Inversion method is equal to SA model at $\alpha < 8.2°$), and the detailed values and the relative errors of the lift coefficients are displayed on Table 1. From the results, we can clearly see that the POD-Inversion data assimilation method can significantly improve the numerical accuracy for the simulation of separated flow, and the relative error of lift coefficients can be reduced from 30% to less than 6% compared with the experimental data.

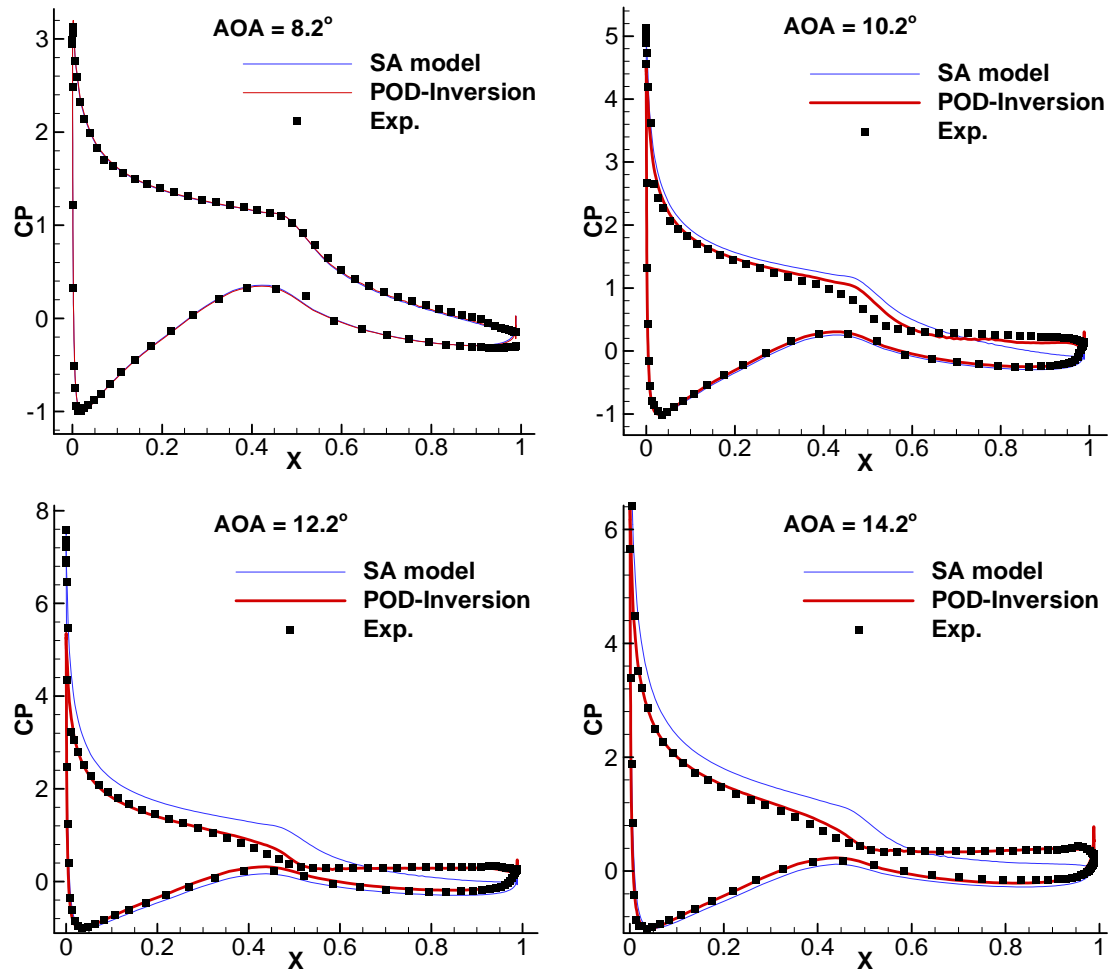

Figure 15 The comparisons of the pressure coefficients for numerical results with the experimental data at different flow states

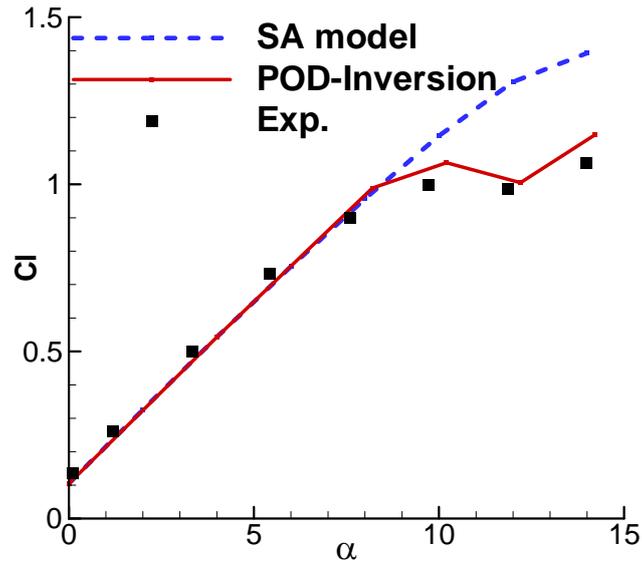

Figure 16 The comparisons of lift coefficients versus angle of attack between experimental data and numerical results

Table 1 The lift coefficients and the relative errors for different method

|  | $\alpha = 8.2°$ | $\alpha = 10.2°$ | $\alpha = 12.2°$ | $\alpha = 14.2°$ |
| --- | --- | --- | --- | --- |
| Experiment | 0.964 | 1.015 | 1.009 | 1.086 |
| SA model | 0.976(1.28%) | 1.163(14.58%) | 1.320(30.82%) | 1.391(28.08%) |
| POD-Inversion | 0.980(1.66%) | 1.064(4.83%) | 1.005(0.40%) | 1.147(5.62%) |

Due to the lack of advanced experimental measures, we can only obtain the pressure coefficients on the airfoil, and no enough friction coefficients are available. Therefore, when we conduct the POD-Inversion method to recover the turbulent flow field, we do not pay attention to the influence of the friction. However, although the pressure coefficients are only used as the optimization target, the whole recovering turbulent flow fields are satisfied with the NS equations. In addition, in condition of separated flow at high angle of attack, the separation point and the size of region also have the key influence on the distributions of friction coefficients. Therefore, we consider the properties of drag can be also augmented simultaneously. Figure 17 displays the distributions of friction coefficients at the angle of 14.2°for POD-Inversion method and SA model, which are also compared with Duraisamy's results. We can see from the Figure 17, the friction coefficients of SA model and POD-Inversion method are both coincided with the reference results. Furthermore, we also compute the lift-drag coefficients at different angle of attack, which is shown in Figure 18. The curve is obviously different between SA model and POD-Inversion method at high angles of attack, but both are in good agreements with those of Duraisamy. It indicates that the turbulent flow fields recovered by POD-Inversion data assimilation method consistent with the physical law, which can accurately compute the separation point and the size of separation region. Although only constrained with the pressure coefficients, the proposed method can both improve the numerical accuracy of pressure and friction coefficients.

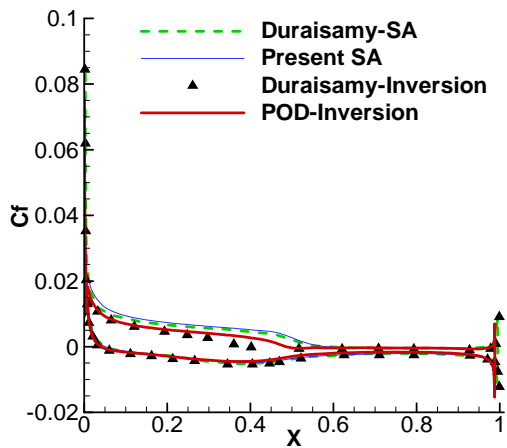 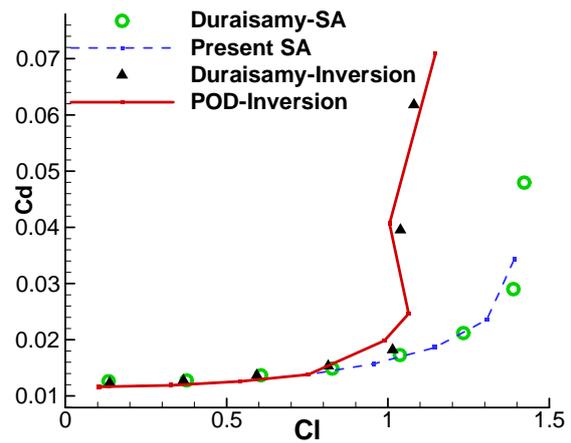

Figure 17 The distributions of friction coefficients for S809 airfoil at $\alpha = 14.2°$ compared with the reference results

Figure 18 The lift-drag coefficients at different angle of attack for S809 airfoil compared with the reference results

## 4. Conclusions

This paper proposes a POD-Inversion data assimilation method for recovering high Reynolds number turbulent flows based on the experimental pressure coefficients. POD technique is introduced to analyze the turbulent eddy viscosity flow fields which are computed by the traditional SA model, and only a few set of main POD modes are remained to reconstruct the turbulent field. It can dramatically reduce the system dimension by using the POD-Inversion method, and achieve the global optimal solutions conveniently. The turbulent eddy viscosity, reconstructed by POD modes and expansion coefficients, is loosely coupled with the standard NS equations solver, and we eventually recover the high-fidelity turbulent flow in condition that the computed pressure coefficients are optimally matched with the experimental data.

The proposed method has been used in two turbulent flows at high Reynolds number, the transonic flow with shock waves around RAE2822 airfoil and the separated flow at high angles of attack around S809 airfoil. In both test cases, we have observed the satisfactory recovering of the high fidelity turbulent fields. The numerical accuracy of pressure coefficients have been significantly improved compared with the traditional SA model, and the relative error of lift coefficients are also reduced from more than 30% to less than 6% referring to the experimental data.

We have also discovered that even though only pressure coefficients are used as the optimization target, the POD-Inversion data assimilation method can both improve the numerical accuracy of lift and drag coefficients. It indicates that the proposed method is a valuable tool to recover high fidelity turbulent flows by a limited amount of flow information. We believe that the POD-Inversion data assimilation method will have great potential in extensive applications.

## Declaration of Competing Interest

There is no conflict of interest for this paper.

## Acknowledgements

This work was supported by the foundation of National Key Laboratory of Science and Technology on Aerodynamic Design and Research (No. 6142201190301), the Chinese National Natural Science Foundation for Young Scholars (No. 11902270), the project of National Numerical Wind-Tunnel (2018-ZT1B01) and the National Natural Science Foundation of China (No. 11572252, No. 91852115).